# Delivering IT as A Utility- A Systematic Review

Inderveer Chana[1] and Tarandeep Kaur[2]

Thapar University, India
[1]inderveer@thapar.edu
[2]tarandeep.kaur@thapar.edu

## ABSTRACT

*Utility Computing has facilitated the creation of new markets that has made it possible to realize the long-held dream of delivering IT as a Utility. Even though utility computing is in its nascent stage today, the proponents of utility computing envisage that it will become a commodity business in the upcoming time and utility service providers will meet all the IT requests of the companies. This paper takes a cross-sectional view at the emergence of utility computing along with different requirements needed to realize utility model. It also surveys the current trends in utility computing highlighting diverse architecture models aligned towards delivering IT as a utility. Different resource management systems for proficient allocation of resources have been listed together with various resource scheduling and pricing strategies used by them. Further, a review of generic key perspectives closely related to the concept of delivering IT as a Utility has been taken citing the contenders for the future enhancements in this technology in the form of Grid and Cloud Computing.*

## 1. INTRODUCTION

Utility Computing is a concept established by John McCarthy who had predicted in the late 1960s that computation may someday be organized as a public utility [1]. The main objective behind Utility computing is to provide the computing resources, such as computational power, storage capacity and services to the users and charging them for their usage. It is a business as well as a technological computing model where the business model envisions that customers pay for the exact usage of the services provided by the service provider and the technological model provides the necessary IT infrastructure to provision the services like utilities [2]. The financial benefits envisaged from metered IT services offered by utility computing aid in achieving better economics.

The initial scenario of high performance computing paradigm consisted of development of supercomputing which was soon replaced by clusters as the supercomputers proved to be incompetent to handle problems in different areas of science, engineering, and business. Though clusters provided high availability, load-balancing and large computation power but they lacked economic considerations which are the characteristic feature of Utility services.

The perception of utility computing began to arise with the significant advancement in IT. It involved transitions from dedicated systems to the shared infrastructure followed by the rise of standardized technological developments. This also included development of real-time management views in the assisted management scenario which further got transformed to actionable infrastructures with an optimized resource management [2]. However, vendors like

IBM, HP, Oracle etc have built large data centers with the view of realizing the utility computing model [3]. In fact, Utility computing has become a popular area of interest in today's commercial and business world. The applicability of utility computing for the potential areas (as discussed in Table 1) is based on general technological and business possibilities in the global market [2].

The objective of this survey is to get a systematic review of the concept of Utility services delivered through the I.T infrastructure. It can help the researchers interested in Utility computing area to carry out future research into it. It highlights the research done towards Utility computing resource scheduling and resource pricing along with the key perspectives of Utility systems.

Table 1: Applications of Utility Computing

| Application Area | Key Services | Pricing Criteria | Major ASP's For Such Services |
|---|---|---|---|
| Web Hosting Service | Internet hosting service | Provide space on a server owned or leased for use by clients. | Olive, Webhostinguk.com |
| E-mail Service | Emailing services | A base price for each new user and variable price according to the size of the mailbox and other functionalities. | Webhostinghub.com, Hostpapa.com, Gmail, Hotmail. |
| Groupware Service | Intentional group processes plus software to support them [4] | A base fee for each user and variable fee calculated for various functions. | Lotus Notes and Microsoft Exchange |
| CRM Service | Organize, automate, and synchronize sales, ma-rketing and customer services [6] | Apart from the base fee, a variable fee is charged according to the functionalities used in accessing CRM services. | Salesforce.com, NetSuite, Salesnet and Siebel CRM OnDemand [5] |
| ERP Service | Integrate internal and external management information | Charging according to the resource usage or time spent online to access the ERP services. | SAP On Demand, Net-Suite hosted for SMBs Salesforce.com, Oracle |
| Storage Service | Storage as a utility service | Charges users using metrics available– megabytes or gigabytes of storage. | EMC or Hewlett-Packard (HP) |
| Computing Service | Pooling the computing resources | Sun's N1 pay-per-use is charged at merely USD 1 per CPU hour. | Sun's N1 [7] |
| Network Service | Network services like WAN, LAN or VPN | Subscription based pricing with base and a variable fee according to the amount of traffic carried or time spent online. | AT&T [8] |
| Data center Service | Virtualizing computing resources | Charging a base fee plus a variable fee according to the total resource usage. [2] | IBM's UMI [9] |

## 2. RESEARCH METHOD

The research method followed in this study is based on finding out relevant research papers from different databases and then framing out the questions that are needed to be addressed.

### 2.1 Research Questions

This review aims at summarizing the present state of the art in utility computing concept by proposing answers to the following questions:

1. What is Utility Computing and which are the key areas where it is used?
2. How to realize the Utility Computing Model?
3. Which service provider can cater to the changing market demands and provide the users with the QoS they require?
4. What services are offered by the service providers and to what extent the services can be easily incorporated into the existing system without posing any technological and business risks ensuring availability, security, economies of scale- factors crucial to the smooth functioning of the client organisation?
5. What pricing policy can be incurred for the resource usage by the service vendors?
6. How service providers can maximise their net revenue ensuring fairness among clients?

### 2.2 Sources of Information

In order to gain a broad knowledge, we searched widely in electronic sources. The databases covered are:

- IEEE eXplore (<ieeexplore.ieee.org>)
- Springer LNCS (<www.springer.com/lncs>)
- ACM Digital Library (<portal.acm.org>)
- Google Scholar

These data sources cover up all the related journals, conferences and workshop proceedings.

### 2.3 Study Selection

The systematic review process started with defining research questions as stated in section 2.1 which was followed by identifying possible search keywords (Table 2) which served as the basis for searching for various research papers. A thorough literature survey was conducted on what actually is the concept of Utility Computing. The official websites of individual service providers were visited to find the kind of utility service provisions available through them. Next step included gaining knowledge about how scheduling of the user requests is carried out and what pricing mechanisms are implemented. Finally, the reference analysis was conducted to ensure that referenced papers were not missed out. Reference analysis is important to avoid missing out certain publications during the keywords-based search using the search engines.

Table 2: Search Keywords and Synonyms

| Keywords | Synonyms |
|---|---|
| Utility Computing | Utility Cloud Computing, Utility Grid Computing |
| Utility Computing Application Service Providers | Utility Computing Service vendors |
| Scheduling in Utility Computing | Resource Scheduling in Utility Computing |

## 3. STEPPING TOWARDS UTILITY COMPUTING

Grid, Cloud, and Service-Oriented Computing are some of the paradigms that are making the delivery of computing as a utility possible [10]. However, for realizing the delivery of IT as a Utility, several key components have been identified as to be important (Figure 1). These components can also be considered as the prerequisites for stepping towards Utility computing.

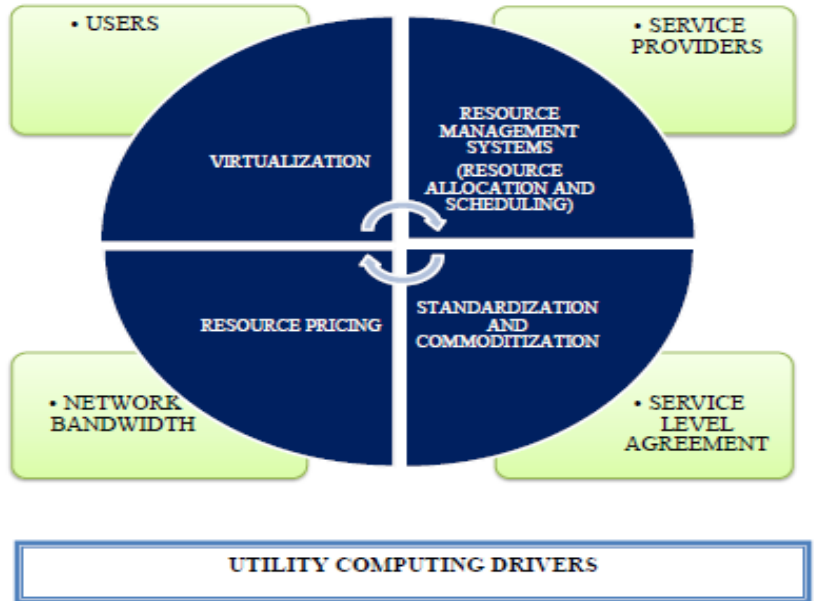

Figure 1: Utility Computing Driving Factors

i.   Application Service Providers: Application service providers control all the necessary resources and deliver application functionality from any location (locally or globally) as a service to the users who no longer have to invest heavily on or maintain their own computing infrastructures. Instead, users can outsource jobs to the service providers and just pay for what they use [10, 11].

ii.  Standardization and Commoditization: Technological standards are vital for the use of utility services. Standardized IT services enable transparent use of services and make a service an ideal candidate for utility computing. The commoditization of hardware and to some extent software such as office suites happens with the increasing acceptance of the open standards [2].

iii. Virtualization: The increasing virtualization of the servers and storage resources has become one of the leading drivers of utility computing [2].Virtualization has enabled the abstraction of computing resources such that a single physical machine is able to function as a set of multiple logical VMs [11]. A key benefit of these VMs is their ability to host multiple operating system environments which are completely isolated from one another on the same physical machine [11, 12, 13, 22].

iv.  Network Bandwidth: The network bandwidth is a critical factor for the successful provisioning of IT services over the network. The arrival of the utility computing has massively increased the traffic over the networks; therefore the network lines must handle the traffic to enable the simultaneous use of the shared applications by the customers [2].

v.   Service Level Agreement (SLA): A SLA can be defined as an explicit statement of expectations and obligations that exist in a business relationship between two organizations: the service providers and the customers [14]. It is actually a formal contract used to guarantee that user's service quality expectation can be achieved. By using SLAs to define service parameters required by the users, the service provider knows how users value their service requests, hence it provides feedback mechanisms to encourage and discourage a service request [10, 15].

vi.   Resource Management System (RMS): The complexity of managing the resources drives the need for an efficient resource management system that must handle crucial tasks such as accepting and rejecting the user requests, starting the components of the accepted requests on private/public resource and migrating components vulnerable for deadline violation from the public resources to the dedicated resources [16]. It helps to maximise the net revenue and minimize the time for execution of an application [17].

vii.   Resource Scheduling: Scheduling the incoming user requests in such a way as to satisfy user QoS and maximize service provider profit is a challenging problem. A scheduler is an important component inside a RMS which discovers the resources according to the requirements of a job (here job refers to a user request), performs the resource allocation and then maps the resources to that job. It has to explicitly account for the amount of data each job needs otherwise certain jobs will occupy the CPU resources for much longer than necessary [18].

viii.   Resource Pricing: The primary function of the resource pricing component is to develop a pricing mechanism that can successfully support economy based utility computing paradigm. Thus, a dynamic and adaptive pricing strategy is required such that the two main user-centric evaluation parameters: QoS satisfaction and provider profitability can be accomplished successfully [2].

# 4. KEY FEATURES OF UTILITY COMPUTING SYSTEMS

The prevalence of Utility computing is a result of advanced research in this field. The key features of Utility computing systems can be:

## 4.1 Utility Computing Architecture Models

Many utility computing architecture models proposed till date consider the services to be offered to the clients according to the QoS constraints imposed in the SLA's while some restrict their area to completing the dynamically changing on-demand requests by the clients. There are certain models dedicated towards achieving maximum utilisation of resources and also preventing certain requests from using the resources for longer periods. Table 3 lists various utility computing architecture models.

Table 3: Various Utility Computing Architecture Models

| Architecture Model | Basis | Key Components | Goal |
|---|---|---|---|
| Service-Oriented Utility Model [10] | Service provisioning model | • User/Broker<br>• Service Request Examiner and Admission Control<br>• SLA Management and Resource Allocation Mechanism<br>• Service Provider | Provides user-driven service management, computational risk management and autonomic resource autonomic. [11] |
| SLA Based Utility Model [19] | Using predefined SLA | • Consumer/Application<br>• SLA aware planning and co-ordination<br>• SLA negotiation and supervision Middleware<br>• Provider | Enabling resource orchestration and co-allocation. |
| Resource-Oriented Utility Model [20] | Peer-to-peer and Grid computing ideas | • Application<br>• Resource pool<br>• Resource addressable network<br>• Resource management unit<br>• Incentive/ Trust management<br>• Security system | Enables commoditization and utilization of the ideal capacity of shared public resources, augumenting the capacity with dedicated resources. |
| Business-Driven Utility Model [21] | Service delivery framework, B2B based architecture | • Service provider<br>• Service aggregator<br>• Service channel maker<br>• Service hoster/ service gateway<br>• Service consumer | Support for diverse provisioning partners, multiple industries and deployments. |
| Market-Oriented Utility Model [22] | Cloud environment | • User/ Broker<br>• SLA resource allocator<br>• Virtual machines<br>• Physical machines | Regulates the supply and demand of resources at market equilibrium achieving high QoS based resource allocation. |
| Model- Based Utility Model [23] | Creating models of IT environment and developing tools that use those models. | • Resource composition<br>• Resource assignment<br>• Capacity management<br>• Resource pool<br>• Service deployment<br>• Operations control | Supports design, deployment and management of arbitrary applications while dealing with frequently competing requirements for resources. |
| Content-Delivery Based Utility Model [24] | Peer of content delivery networks (CDNs). | • Peering agent<br>• Content providers<br>• Primary CDN's<br>• Request queuing and scheduling<br>• Resource negotiation<br>• SLA monitoring<br>• Service discovery | Focuses on optimizing the content delivery to internet end-users from multiple, geographically distributed replica servers. |

## 4.2 Utility Based Resource Management Systems

The efficiency of the resource usage is maximized through the automated allocation of IT resources by the resource management unit [2]. The RMS manages large pools of resources that are available as a service to the customers. This pooling of resources reduces the operating costs of the IT infrastructure and thus provides economies of scale. Different resource management systems developed so far handle resource allocation efficiently in the utility systems and have been listed in the Table 4 below.

Table 4: Different Types of Utility Resource Management Systems

| Resource Management Systems | Key Features | Examples |
|---|---|---|
| Market Based RMS [12,25,26] | Regulate the supply and demand of resources to achieve market equilibrium, providing feedback in terms of economic incentives for both users and providers, and promoting QoS. [26] | • Cluster-On-Demand<br>• Enhanced MOSIX<br>• REXEC<br>• GridBus<br>• Nimrod/G |
| Enterprise Based RMS [27,28] | Economy-based Resource Allocations. | • Enterprise<br>• Spawn<br>• POPCORN |
| SLA Based RMS [29] | Encompass both customer-driven service management and computational risk management. | • LibraSLA |
| Cluster RMS [30] | Aim at providing high-performance, high-throughput and high-availability computing services. | • Condor<br>• Maui<br>• LoadLeveler |

## 4.3 Utility Scheduling Strategies

Scheduling in the utility computing systems is an important and difficult task. A large number of user requests keep on entering the utility system from time to time. Most of these requests that enter the system need to access some data stored on local disk or remote storage, while others need to access very large amounts of data. Whenever a request finishes it brings some advantage to the system in the form of utility and therefore the aim is to find an optimal scheduling policy that maximizes the average utility per unit of time obtained from all the successfully finished requests [23]. Different resource scheduling schemes have been proposed so far. The key characteristics of these schemes are mentioned below along with a list their corresponding example systems (Table 5).

- First in First out Scheduling Strategy: In this strategy, tasks are assigned in the order they arrive in a queue to an arbitrary resource which meets the task's requirements. This strategy implies that the priority of a task is equal to the time it has spent waiting in the queue (i.e. current time minus submission time) [15].

- Market Based Scheduling Algorithm: It is a business centric approach based on user valuation of QoS parameters as specified by the users and profitability which is the aim of the service providers. It involves using Economy-based admission control system and also uses a dynamic and adaptive pricing mechanism [30].

- Adaptive Request Scheduling Strategy: This scheme handles an adaptive request, the one that can be modified in terms of not only the number of resources and usage time, but also whose starting and completion time are not fixed when the request is scheduled. Here, the resources are defined as the cluster nodes managed by RMSs that contain schedulers responsible for receiving the user requests and placing them into a waiting queue according to the available timeslots [31].

- Multiple QoS based Resource Scheduling Strategy: The objective of this scheduling scheme is to maximize the global utility by scheduling the finite resources. Each of user's diverse requirements is modeled as a quality QoS dimension, associated with each QoS dimension is a utility function that defines the benefit perceived by a user with respect to QoS choices in that dimension [32].

- Round Robin Scheduling: This involves allocating different time slices to each user request and executing them according to each time slot allocated to them.

- Budget and Deadline Constrained Scheduling: This scheduling technique requires economy driven deadline and budget constrained (DBC) scheduling algorithms for allocating resources to application jobs in such a way that the users' requirements of minimized time and cost are met [33].

Table 5: Scheduling Techniques Used in Various Schedulers

| Scheduling Startegy | Example Systems |
| --- | --- |
| FCFS | GridWay, Condor-G |
| Market Oriented Scheduling | Tyccon, SPAWN, Bellagio, SHARP, Libra |
| QoS Based Scheduling | Nimrod- G, POPCORN |
| Round Robin Scheduling | REXEC |
| Budget and Deadline Constrained Scheduling | REXEC, AppLeS, NetSolve |

## 4.4 Utility Computing Pricing Mechanisms

The utility services involve pricing based on multiple, measured tiers such as response time, availability, throughput etc [2]. Thus, an effective pricing mechanism to support economy-based resource management and allocation in utility systems is required [30]. Different payment models are being used by the service providers to charge the customers. Some of these models along with the different resource management systems based on these models are:

i. Flat Rate Model: A flat fee, also referred to as a flat rate or a linear rate, refers to a pricing structure that charges a single fixed fee for a service, regardless of usage. For Internet service providers, flat rate is access to the Internet at all hours and days of the year (linear rate) at a fixed and cheap tariff [34].

ii. Tiered Model: Service providers are increasingly selling "tiered" contracts, which offer utility services to wholesale customers in bundles, at rates based on the cost of the links that the traffic in the bundle is traversing. An approach is by increasing block rates (hereafter, IBR or tiered pricing), where individuals pay a low rate for an initial consumption block and a higher rate as they increase use beyond that block [35].

iii. Subscription Model: The subscription model is a business model in which the customers pay a subscription price to have access to the products/services. Industries that use this model include cable television, satellite television providers, telephone companies, cell phone companies, internet providers, software providers [36].

iv. Metered Services: Metered services involve measuring the amount of consumption of the resources used by the users and then billing them on the basis of the meter reading. This model prevents users from paying extra as in case of fixed price model.

v. Pay as you go Model: It involves charging customers based on their usage and consumption of a service. IBM offers services based on pay-as-you-go model. It is also called as "Pricing based on consumption".

vi. Standing Charges Scheme: Tariffs with a standing charge involve payments at flat rate charges for the supply as well as for the amount of power usage.

On a broader scale, pricing models can be classified as fixed price and variable price models. Table 6 below illustrates these models along with the list of systems based on these models [37].

Table 6: Utility Pricing Models

| Model | Characteristics | Systems using the model |
|---|---|---|
| Fixed Price Models | • Fixed pricing is the simplest pricing option, where the billable item has a fixed periodic cost. | • Libra<br>• Tycoon<br>• Bellagio<br>• Mirage |
| Variable Price Models | • Variable pricing by resource consumption involves billing the consumer for the actual amount of atomic units of the billable item that were used.<br>• Variable pricing by time involves billing the consumer based on how long the billable item was used. | • FirstPrice<br>• FirstReward<br>• FirstProfit<br>• FirstOppurtunity |

The major objective of every service provider is to earn higher profits. The need to realize higher net revenue compels major steps to be taken by the service providers to improve their standards of services. The graph (Figure 2) below gives a list of the top companies in the IT services industry, ranked by their Total revenues for the year 2011. The list consists of a large number of system integrators, software implementation companies, communication services suppliers, outsourcing partners and data center specialists [38].

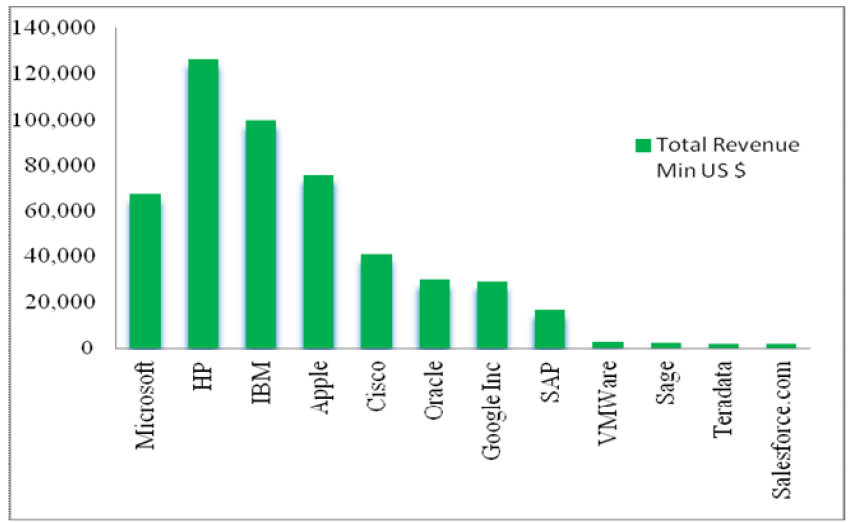

Figure 2: Top software companies ranked by the Total Services Revenue, 2011

# 5. DATA ANALYSIS AND RESULTS

This section explains the findings of the survey. An analysis of different companies active in the market of utility computing was carried out. It included companies such as Google, Amazon Web Services, IBM, Hewlett Packard (HP), Flexiscale etc. The study included finding the services offered by them and predicting the profit shares of the different companies.

## 5.1 Top Utility Service Providers

Table 7 below gives the top Utility computing service providers existing in the market which can satisfy the user needs and are engaged in providing different utility services such as storage facility, web hosting facility etc. The column "Offerings" highlights the quality of services offered by them along with the list of their solutions available and their client organisations.

Table 7: Top Utility Computing Service Providers

| Service Providers | Offerings | Solutions | Clients |
|---|---|---|---|
| IBM Smart Cloud [39] | <ul><li>High-performance</li><li>Scalability</li><li>Security-rich</li><li>Simplified restore capabilities</li><li>Web-based security services</li></ul> | <ul><li>Smart Business Storage Cloud.</li><li>IBM SmartCloud Archive</li><li>IBM SmartCloud managed backup services</li><li>IBM SmartCloud Virtualized Server Recovery</li><li>IBM Managed Security Services Management</li></ul> | <ul><li>SAP</li><li>Oracle</li><li>Lotus</li><li>PeopleSoft</li></ul> |
| Google [40] | <ul><li>Scalability</li><li>Cost saving</li><li>High Speed</li></ul> | <ul><li>Google App Engine</li><li>Google Cloud Storage</li></ul> | <ul><li>Gigya</li><li>Mimiboard</li><li>Hudora</li><li>Ubisoft</li><li>PocketGems</li></ul> |
| FlexiScale [41] | <ul><li>Flexibility</li><li>Scalability</li><li>High QoS</li><li>Reliability</li><li>Compatibility</li><li>Fully virtualized storage.</li></ul> | <ul><li>Flexiant Cloud Orchestrator</li><li>Reseller Hosting: White Label FlexiScale</li><li>FlexiScale 2.0: Flexible & Scalable Public Cloud Hosting</li></ul> | <ul><li>Logica</li><li>Yahoo</li><li>Oracle</li><li>Telecom Italia</li><li>ERISS</li></ul> |
| HP Converged Cloud [42,43] | <ul><li>Flexibility</li><li>Simplicity</li><li>Scalability</li><li>High Speed</li><li>Reduced cost and risks</li><li>Efficiency</li><li>Enhanced disaster</li></ul> | <ul><li>Enterprise Cloud Services</li><li>Cloud Consulting Services</li><li>Cloud Education Services</li><li>Cloud Education Services</li><li>Data Center Services</li></ul> | <ul><li>CloudAgile</li><li>Avaya</li><li>Citrix</li><li>ExpertOne</li><li>PartnerOne</li></ul> |

| | recovery | | |
|---|---|---|---|
| Amazon EC2 (Elastic Compute Cloud) [44] | • Scalability<br>• Pay-as-you Go<br>• Flexibility<br>• Reliability<br>• Secure<br>• Inexpensive<br>• Elasticity | • Amazon Elastic Block Store<br>• Amazon Virtual Private Cloud<br>• Amazon CloudWatch | • Zoomii<br>• Morph<br>• Washington Post<br>• Many Facebook applications |
| Amazon S3 [45] | • Highly scalable<br>• Data Durability and Reliability<br>• Pay-as-you Go<br>• Security<br>• Fast<br>• Inexpensive | • Amazon Glacier<br>• Amazon CloudFront<br>• Amazon EBS | • DragonDisk<br>• Posterous<br>• Tumblr<br>• Formspring |

## 5.2 Benefit Levels of Top Utility Computing Service Providers

Several vendors such as Amazon, IBM, Microsoft, Google, HP offer utility products and are now the leading providers of utility services. The chart (Figure 3) below displays the top utility cloud computing service providers in the world for 2011 with the percentage of benefit they incurred on using utility cloud computing [46].

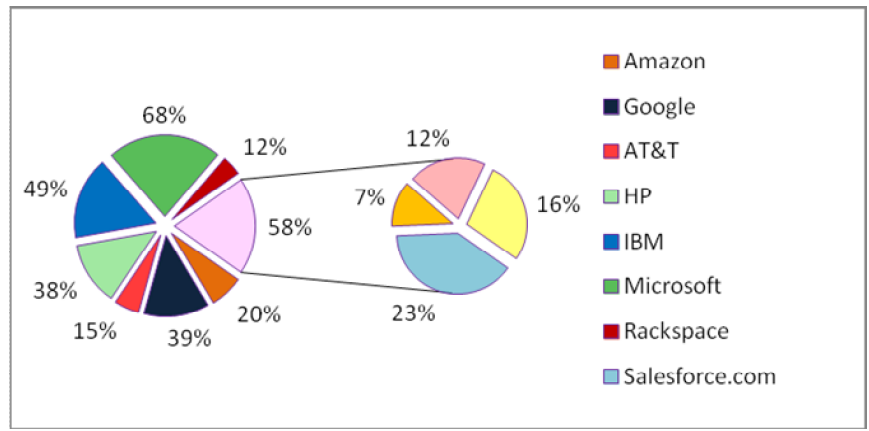

Figure 3: Beneficiary percentage levels of top utility service providers, 2011

# 6. POSSIBLE SOLUTIONS

The major goal of Utility computing is to provide the services on demand speedily (i.e. as required by the end-user organization). This section deals with pointing out the possible solutions to actually realize the idea of utility computing either via a Grid (geographically dispersed, interconnected, heterogeneous and dynamically configurable computing resources) or a cloud (a form of distributed computing whereby a "super and virtual computer" is composed of a cluster of networked, loosely-coupled computers, acting in concert to perform very large tasks) [5]. It outlines different benefits of Grid and Cloud that make them ideal for being identified as true utility provision paradigms.

## 6.1 Delivering Utility Services through Grid

Delivering the Utility services was formerly unrealistic or impractical due to globally dispersed nature of vital resources. But in recent years, Grid computing has emerged as a fast developing technology with the capability to surmount the bottleneck of inability to utilize the capability of physically dispersed resources. Grid computing is a promising concept to enhance the effectiveness of existing computing systems and to cut down on IT expenditures by providing dynamic access to computer resources across geographical and organizational boundaries [47].
Grid is a type of parallel and distributed system that enables the sharing, selection, and aggregation of geographically distributed autonomous resources dynamically at runtime depending on their availability, capability, performance, cost, and users QoS requirements [48] [49]. Grid Computing is the ability to gain access to applications and data, processing power, storage capacity and a vast array of other computing resources over the Internet using a set of open standards and protocols [50] [51]. Grids offer following services as utilities:

- Computational power of globally distributed computers through Computational Grid.
- Organisation of data for data access, integration and processing through dispersed data storehouses via Data Grid.
- Access to remote applications, modules and libraries via Application Service Provisioning (ASP) Grid.
- Interaction and joint visualization between participants via Interaction Grid.
- Knowledge management, particularly knowledge acquirement, processing and management through Knowledge Grid.
- Other Grid services such as computational power, data and services to end-users as IT utilities on a subscription basis via Utility Grid. For example, Sun's N1 Grid Engine offers compute power at USD 1 per CPU hour [7].

Grids offer a number of benefits which make them favorable for delivering Utility services which include:

- Coordinated resource sharing and problem solving through dynamic, multi-institutional virtual organizations [52].
- Delivery of transparent and on-demand access to distributed and heterogeneous resources.
- Seamless computing power exploiting the under-utilized or unused resources.
- Provisioning of extra resources to solve problems that were previously unsolvable due to the lack of resources [50] [53].
- Enhance the productivity justifying the IT capital investments [50] [53].

After years of development in the Grid computing under several projects, it has been observed to be an effective way for managing the resources, satisfying the QoS constraints. Some of the projects in Grid computing are briefed in Table 8:

Table 8: Analysis of Some of the Existing Grid Systems

| Project | Grid Systems | Findings |
|---------|--------------|----------|
| P1 | Nimrod [54,55,56,57] | 1. Computational and Service Grid<br>2. Computational economy driven architecture<br>3. Soft-deadline and budget based scheduling |
| P2 | Condor-G [54,57,58,59] | 1. Computational Grid<br>2. Cycle Stealing Technology<br>3. Ad hoc failure detection Mechanisms |

| P3 | SOGCA [60] | 1. Service-oriented grid architecture<br>2. Manage the service provider dynamically |
| P4 | Globus [48] | 1. Computational Grid<br>2. Open source Grid services Architecture |
| P5 | Sun N1 Grid Engine [7,61] | 1. Commercial version of Sun Grid Engine<br>2. Powerful policy-based resource allocation |
| P6 | GridBus [62] | 1. Computational and Data Grid<br>2. End-to-End QoS<br>3. Manages distributed computational, data, and application services |

Some other Grid projects are the NSFs National Technology Grid, NASAs Information Power Grid, GriPhyN, NEESGrid, Particle Physics Data Grid and the European Data Grid [48]. All these projects are working towards accomplishing the Grid computing paradigm successfully. The comparison of some of the existing Grid systems based on certain metrics is shown in Table 9:

Table 9: Comparison of Existing Grid Systems Based On Various Metrics

| METRICS | P1 | P2 | P3 | P4 | P5 | P6 |
|---|---|---|---|---|---|---|
| SCALABILITY | NO | NO | YES | NO | YES | YES |
| FAULT TOLERANCE | NO | YES | YES | YES | NO | YES |
| QoS SUPPORT | YES | NO | NO | YES | YES | YES |
| ECONOMIC CONSIDERATIONS | YES | YES | NO | YES | YES | YES |
| FLEXIBILITY | YES | YES | YES | YES | YES | YES |
| SECURITY | NO | NO | NO | YES | YES | YES |
| PERFORMANCE AND EFFICIENCY | YES | YES | NO | YES | YES | YES |

However, there are major hurdles in the success of Grid concept. One underlying challenge in Grid computing is the effective coordination of resource sharing in dynamic, multi-distributed environment [15]. The task of providing a powerful and robust platform that can make the resources accessible to the users easily without any deep technical knowledge is a complex job. On the other hand, the services such as job scheduling, load balancing, resource discovery and allocation pose intricate management risks [63, 64]. Other critical success factors include ever changing market demands and design of efficient resource scheduling algorithms. The study is also being conducted on developing an energy efficient Grid datacenter that aims at using various energy conservation strategies to efficiently allocate resources.

## 6.2 Delivering Utility Services through Cloud

Cloud computing presents a new model for IT service delivery and it typically involves over-a-network, on-demand, self-service access utilising pools of often virtualized resources [65]. This model comes into focus only when there is a thought about what IT always needs: a way to increase capacity or add capabilities on the fly without investing in new infrastructure, training the new personnel, or licensing a new software [65].The cloud computing architecture comprises of cloud services (measured services) delivered by cloud service providers to cloud consumers over a networked infrastructure [65].Cloud can offer software (software-as-a-service), hardware (infrastructure-as-a-service) or technology tools (platform-as-a-service) as services and can be deployed in three different models:

- Public cloud, run by a third company, (e.g. Google, Amazon, and Microsoft) and provides cloud services, such as storage systems and network.
- Private cloud is built by one client and offers data security and high QoS.
- Hybrid cloud is a combination of both public and private cloud models. It provides on-demand external provision of hardware and networking facilities [66].

Such cloud computing offerings are governed by contractual agreements (SLA) that specify consumer requirements and the provider's commitment to them [13,62].The recently emerged cloud computing paradigm appears to be the most promising one to leverage and build on the developments from other paradigms [13,67]. Cloud computing provides large number of benefits. The key cloud benefits include [68]:

- Independence from hosting infrastructure provided by the data centers monitored and maintained around the clock by service providers.
- Reduced cost as highly robust infrastructure is typically provided by a third-party and does not need to be purchased by users.
- Device and location independence enables users to access systems regardless of their location or what device they are using.
- Virtualization technology allows servers and storage devices to be shared and utilization be increased.
- Scalability and Elasticity via dynamic ("on-demand") provisioning of resources on a fine-grained, self-service basis, without users having to engineer for peak loads.
- Multitenancy enables sharing of resources and costs across a large pool of users thus allowing for:
- Centralization of infrastructure in locations with lower costs (such as real estate, electricity, etc.)
- Utilisation and efficiency improvements for systems that are often only 10–20% utilised.

This form of utility computing is getting new life from Amazon.com, Sun, IBM, and others who now offer storage and virtual servers that IT users can access on demand. Some of the Cloud projects are listed in Table 10 [69]. These projects are working towards gaining momentum in the field of Cloud computing.

Table 10: Analysis of Some of the Existing Cloud Systems

| Project | Cloud Systems | Findings |
|---|---|---|
| P1 | Amazon EC2 [44] | 1. Part of Amazon Web Services.<br>2. IaaS Cloud model<br>3. Offers resizable compute capacity in the cloud<br>4. Provides elasticity to the users to create, launch, and terminate server instances as needed |
| P2 | Amazon S3 [45] | 1. An online storage web service<br>2. IaaS Cloud model<br>3. Highly durable storage infrastructure |

| P3 | Google App Engine [70,71] | 1. PaaS Cloud model<br>2. Google's geo-distributed architecture<br>3. Automatic scaling and load balancing |
|----|----|----|
| P4 | SUN Cloud [71,72,73] | 1. PaaS Cloud model<br>2. On-demand Cloud computing service<br>3. Provides multiple hardware architectures |
| P5 | MapReduce [69,74] | 1. Introduced by Google<br>2. Performs map and reduction operations in parallel |
| P6 | Microsoft Azure [71,75,76] | 1. PaaS Cloud model<br>2. A simple, comprehensive, and powerful platform for the creation of web applications and services. |

The comparison of some of the existing Cloud systems based on certain metrics is shown in Table 11:

Table 11: Comparison of Existing Cloud Systems Based On Various Metrics

| METRICS | P1 | P2 | P3 | P4 | P5 | P6 |
|---------|-----|-----|-----|-----|-----|-----|
| SCALABILITY | YES | YES | YES | YES | YES | YES |
| FAULT TOLERANCE | YES | YES | YES | YES | YES | YES |
| ECONOMIC CONSIDERATIONS | YES | YES | YES | YES | NO | YES |
| FLEXIBILITY | YES | YES | YES | YES | NO | YES |
| SECURITY | YES | YES | YES | YES | NO | YES |
| PERFORMANCE AND EFFICIENCY | YES | YES | YES | YES | NO | YES |

Although Cloud computing has been widely adopted by the industry, there are many exigent issues like performance, security, energy efficiency, interoperability, virtual machine migration, server consolidation, load balancing etc which have not been fully addressed [77,78,79].

## 6.3 Delivering Utility through Data Centers

Preparing for utility computing requires building an architecture that helps any service provider firm to define its operations in terms of technology and process components. At an enterprise level this involves defining what technologies will be shared across all business units and what business processes can be replicated globally [80].

It is anticipated that Grids can provide significant benefits in enterprise environments. Specifically, Grids can enable adaptive resource sharing and collaboration within enterprises. Correspondingly, Utility data centers (UDC) offer an adaptive resource provisioning system targeted at enterprise applications. The UDC and Grid jointly can offer complementary sets of technologies making the UDC an effective power station for an enterprise Grid [81]. Therefore, a hybrid approach can be followed that involves bringing together the Grid technology benefits and the data center fineness to accomplish high performance utility systems.

# 7. CONCLUSION AND FUTURE WORK

This review paper has addressed resource scheduling, resource pricing, and other generic utility computing perspectives. Although the utility computing environment has an automated resource management unit but still it needs further research to handle the resource management issue efficiently. Some of the other under-looked areas include security, energy-efficiency, QoS satisfaction, availability, reliability etc. As the security measures are more relevant in the virtualized environment, where customers are not aware about the location of data and applications, security mechanisms are needed to ensure that unauthorized access to the data and applications can be prevented. Moreover, to appropriately assure desired QoS, resource accessibility and efficient allocation of resources are needed. Future research can be diverted towards analysing potential impact of energy saving strategies. The main sources that consume higher energy and the significant trade-offs between performance, QoS and energy efficiency can be identified.

## ACKNOWLEDGEMENTS

Acknowledgement: This research was supported by AICTE under Research Promotion Scheme Ref. No.8023/RID/RPS-151 (pvt)/2011-2012. Authors' addresses: Dr. Inderveer Chana, Computer Science and Engineering Department, Thapar University, Punjab, Email: inderveer@thapar.edu ; Tarandeep Kaur, Computer Science and Engineering Department, Thapar University, Punjab, Zip Code-147001, Email: tarandeep.kaur@thapar.edu ;

**Authors**

**Dr. Inderveer Chana** completed her B.Tech in Computer Science (1997) and M.E. in Software Engineering from TIET (2002) and Ph.D. in Grid Computing from Thapar University, Patiala (2009) and has over fifteen years of teaching and research experience. She is working as Associate Professor in Computer Science and Engineering Department, Thapar University, Patiala. Her research interests include Grid computing, Cloud computing and resource management challenges in Grids and Clouds. She has over 60 publications in International Journals and Conferences of repute. More than 26 Masters have been completed so far under her supervision and is currently supervising 7 Doctoral candidates in the area of Grid and Cloud Computing.

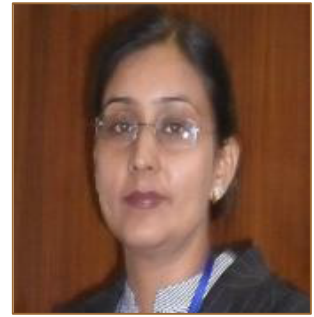

**Tarandeep Kaur** received her Bachelor's Degree (Computer Applications) in 2008 and Master's Degree (MCA) in 2010 from Guru Nanak Dev University, Amritsar, Punjab, India. At present, she is a Ph.D. candidate in Computer Science and Engineering Department at Thapar University, India. Her research interests lie in Cloud Computing, Resource Scheduling, Energy Efficiency and Resource utilization strategies

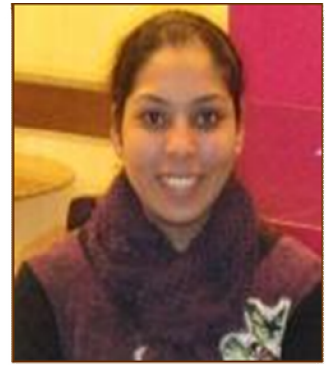